\documentstyle[epsf,twocolumn,aps,prl]{revtex}

\newcommand{\beq}{\begin{equation}}
\newcommand{\eeq}{\end{equation}}

\newcommand{\greaterthanorabout}
           {\mathrel{\raise.3ex\hbox{$>$\kern-.75em\lower1ex\hbox{$\sim$}}}}
\newcommand{\vol}[1]{{\bf #1}}

\begin{document}
\author{Alexei V. Tkachenko$^*$ and Vachtang Putkaradze$^+$\\
$^*$ {\it Bell Labs, Lucent Technologies, 600-700 Mountain Ave. Murray Hill NJ 07974}\\
$^+$ {\it Dept. of Mathematics and Statistics, University of New Mexico, Albuquerque NM 87131-1141}\\
 $^*$ $^+$ Previous address:{ \it The James Franck Institute, The University of Chicago, Illinois 60637} }
\date{to be submitted to {\sl Phys. Rev. Lett.}}
\title{\bf  Mesoscopic Physics of Granular Flows}
\maketitle
  
\begin{abstract}
We present a description of granular dynamics based on the idea of 
differentiation between fluid and solid components.  First, we construct  a 
model of completely fluidized phase. Then we  discuss  a shear surface  motion 
on the boundary of the  bulk solid phase, induced by a moving wall. Our results 
include the thickness  dependence of the velocity of a  developed avalanche,   
density and velocity profiles in granular  Couette flow experiment.


\end{abstract}
  
\bigskip

Dynamics  of granular materials have recently attracted substantial 
attention of physicists\cite{review}.
Despite the numerous experimental and numerical results, 
there is still very little established theoretical  framework for  description 
of granular flows. One of the dominant approaches is based on the 
principles of the kinetic theory of liquids\cite{kinetic}.  Within this 
paradigm, 
it is   believed that the dissipation in  a  granular system does 
  not alter the basic structure of conventional hydrodynamics: the motion is  
expressed in terms of local  velocity, density and (granular) temperature. 
 The dissipation can be accounted for by 
  adding an appropriate dissipative  term to  
the energy and momentum balance equation.
 Below, 
we argue that such an approach can hardly be valid for many granular systems, and   propose an alternative 
approach to  granular dynamics.

 Implicitly, kinetic approach  assumes 
that energy and momentum in the system is transfered through the 
collisions between particles. In the case of hard--core interactions, 
these collision events take essentially no time compared to the free 
flight  of the grains. However, this picture does not survive 
in a general case of a dense dissipative granular system. It is confronted 
by the numerically-discovered phenomenon known as {\em inelastic 
collapse}\cite{Goldhirsh.Zanetti}:  
if we start with an arbitrary distribution of grain velocities, the system
of inelastic grains 
undergoes infinite number of collisions within a finite time and a large 
part of it   virtually stops and creates dense clusters.

 If  the
 system were sheared, the collapsed structures would provide a skeleton  
for the solid--like momentum transfer trough inter--grain mechanical 
contacts. This mechanism  is supported by direct observation of  force--
carrying mesoscopic structures in the sheared granular system\cite{beringer,gauss}. 
These string--like structures closely resemble the force chains responsible 
for stress transmission through static granular material\cite{chains}, and  they are  
obviously inconsistent with the collisional  picture for the momentum 
transfer.   Our argument does not rule out the applicability of the 
kinetic theory    to, for example, 
 sufficiently  dilute granular  systems with weak 
 dissipation, or to granular materials subjected to the bulk pumping of
 energy.

In this letter, we are interested in physics of steady  granular 
flow in the limit of high  dissipation. As it was pointed out, 
in this case the fluid-like motion  appears to be consistent 
with solid--like mechanism for  stress transmission. 
The fundamental reason for this coexistence is that even if 
the contact network becomes  percolating and can transfer the 
stress across the system, its connectivity may still be insufficient 
for mechanical stability (i.e. it may be  below the rigidity percolation 
threshold). Following this observation, we can  distinguish between 
{\em fluid} and {\em solid} states of granular matter, based on its
 local connectivity.
 If the connectivity is high, the granular matter will be 
``jammed'': a bulk of particles is only able to move as a solid, no motion 
of granules with respect to each other is possible.  If, on the other hand, 
 the connectivity of the network is low,   grains can move with respect 
to each other.

 Thus, we can introduce a local ``order parameter'' $\Psi$, as a local fraction of 
particles belonging to a given solid cluster. The motion of particle belonging to the same 
solid cluster is {\em coherent}, which makes   the $(\Psi=1)$--phase  similar to 
more conventional correlated states  known in condensed matter physics, such as 
superconducting or superfluid  condensates. In  the following, it will be  more 
convenient for us to use  mobility parameter $\phi\equiv 1-\Psi$, which is the 
local fraction of the fluid component. 
Similar ideas of characterization of the granular system with the amount of 
fluidized component have been previously successfully used within BCRE model of 
surface avalanches \cite{BCRE}. The distinctive feature of our approach is that 
we are interested in the behavior of the system on the length scale of several 
beads, rather than in its macroscopic phenomenology.

 We limit ourselves to the generic case of planar geometry, 
in which all principle variables, such as time--averaged velocity or the 
 order parameter, vary only  in the direction $z$ normal to that of  motion ($x$ 
 is directed along the flow).
We shall consider two examples: flow of granules down an inclined plane and 
a shear flow  of granular matter  in Couette--type  experiment. In the first 
case, the motion occurs under a given  local  stress, so 
 the principle equations will include forces.
In the second case, the velocity on the inner wall is imposed, which is a 
kinematic condition, so the equations are also kinematic.  

\begin{figure}
 \epsfxsize=\hsize $$\epsfbox{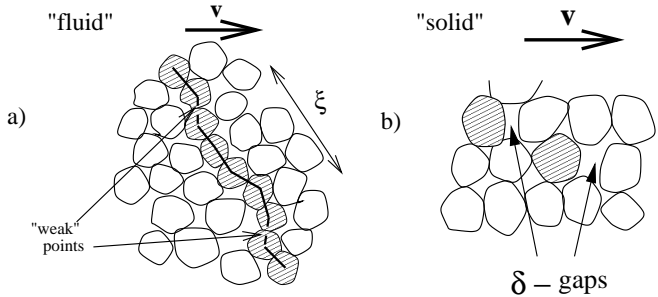}$$
\caption{Illustration to our discussion of flow mechanisms  in ``fluid'' (a) and ``solid'' (b) phases.}
\label{avalanche}
\end{figure}

{\bf Completely fluidized case: $\phi = 1$.}  
We consider the motion of the granular material down an inclined plane, whose 
tilt is large enough for  the system to be completely fluidized. 
In this state, the force chains transmitting an external 
stress are unstable with respect  to  {\em buckling}. In other words, 
the difference between a static force chain and  one in the flow is 
that the latter has "weak bonds"  to be broken  due to external forcing (see Figure \ref{avalanche},a).

The network  rearrangement associated with any of the buckling modes   takes 
a finite time, which can be estimated by noticing that  if $f$ is the typical
 force transmitted by a force chain, the unbalanced torque at a  particle of
 diameter $d$ should be of order of $fd$. Over the buckling time, this torque 
should rotate the particle (of mass $m$)  by an angle of order of unity, until
 a new contact is created. If $\tau_b$ is the buckling time, this gives 
 $m d^2/\tau_b^2 \simeq f d$ or 

\begin{equation} 
\label{taub}  
 \tau_b \simeq \sqrt{\frac{md}{f}}.  
\end{equation} 

The above estimate yields the time scale over which a single "weak" 
bond is destroyed. In a steady flow there is certain   equilibrium fraction 
of the "weak" bonds (their number  is equal to the number of independent ``zero mode'' deformations  permitted by the contact network).  This fraction  can be expressed as $\alpha \equiv d/\xi$, where 
$\xi$ is   the typical length of a stable chain segment between 
two unstable ("weak")  points. Within our model, these stable segments 
 move like a  solid, while the shear is concentrated near the "weak" bonds. 
Therefore, $\xi$ can be interpreted as  {\it  coherence length} associated
 with the order parameter  parameter $\Psi$.   
The shear rate $\partial_z  v$, is the time
 scale over which any particle will change its neighbors. We can estimate this 
rate as $\alpha/\tau$. Now the shear rate can be related to the local
 stress by: $\partial_z v \simeq \alpha \sqrt{\sigma d/m}$, where $\sigma$ is
 the typical value of the stress. In the spirit of recent ideas on the
 constitutive law in  granular solids \cite{Cates,AT.TW}, 
we assume that the stress tensor
 is completely determined   by its  ${\sigma}_{zz}$ and  ${\sigma}_{zx}$ components (given that ${\sigma}_{zy}=0$).
Now, we can  rewrite the earlier 
result 
in the following form:
\beq
\label{dzv}
\left|\partial_z {v}\right| \partial_z {v}=  
\frac{{\sigma}_{zx}d^3}{m \xi \left(\sigma_{zx}/\sigma_{zz}\right)}.
\eeq 
 
This equation of velocity has to be accompanied by an equation for 
the parameter $\alpha$ or, equivalently, coherence length $\xi$. 
It is a natural assumption that $\xi$  is a decreasing function of 
the ratio of the tangential and normal stresses.
 In fact, as this stress ratio 
decreases, the number of free modes  goes to zero, and  $\xi$ should diverge at
 certain critical point, $\sigma_{zx}/\sigma_{zz}= \mu_0$.
  After this jamming transition,  the  granular system becomes 
solid, i.e. $\alpha=0$, and $\xi=\infty$.

If we start with the solid phase at the jamming point  $\mu_0$ and gradually
 increase the ratio $\sigma_{zx}/\sigma_{zz}$, e.g. by changing the slope
 of the free interface with respect to the gravity direction, the solid will
 remain stable until another critical point, $\mu_1>\mu_0$. 
This means that $\xi$ is a two-branch function of the stress ratio: between   
 $\mu_0$ and $\mu_1$,  
$\xi= \infty$ for $\phi=0$, and $\xi$ is finite for $\phi=1$. 
This non-uniqueness and the associated hysteresis, result in the 
 stick-slip response to certain class of  external driving, observed
 experimentally\cite{Gollub.exper}. 

We now apply (\ref{dzv}) 
to the granular flow induced by an external bulk force (e. g. gravitational),
 tilted by  angle $\theta$ with respect to   the $z-$direction. 
The breakage of bonds is dependent on the stress ratio 
$\sigma_{zx}/\sigma_{zz}$, which remains constant
 everywhere and equals to $\tan \theta$. 
 Let the height of the moving granular layer  be $h$, with $z=0$ corresponding
 to its bottom and $z=h$ being its top.  The component 
$\sigma_{zx}$  of the stress is
 essentially given by the "hydrostatic" pressure: 
$\sigma_{zx}=\rho g (h-z) \sin \theta$. Thus, (\ref{dzv}) 
 results in the following velocity field :
\beq 
\label{velip} 
 v(z)\simeq\frac{\sqrt{g} 
\sin(\theta)}{\xi(\tan\theta)}\left(h^{3/2}-(h-z)^{3/2}\right)
\eeq
We conclude that the typical velocity of a thick developed avalanche scales
 as  $h^{3/2}$ with its height, and that the dependencies 
of the average flow velocity on the layer height  collapse onto 
 a single  master curve 
\beq
\label{master}
v\simeq \sqrt{g}h^{3/2}/\xi, 
\eeq
for a variety of physical  parameters (i.e., slope, friction, etc). 
The scaling law $v\sim h^{3/2}$ have  indeed been
  observed   experimentally by Azanza {\em et al} \cite{Azanza}. Moreover, 
Pouliquen \cite{Pouliquen} in recent study  has  demonstrated the validity of 
above master relationship, Eq(\ref{master}).  
In the experiment, 
 the fundamental length scale   $\xi$ reveals itself as the minimal
 layer thickness at which the material  flows for a given slope, which
 is completely  consistent with our interpretation of $\xi$. 
The experiments also support our observation that  $\xi$ diverges 
as the  slope approaches the  critical value, $\tan^{-1}\mu_0$.

{\bf Partially--fluidized case ($\phi\simeq 0$)}. 
Our further discussion is devoted to  the shear dynamics of the  solid phase 
perturbed  by a moving wall, as in Couette flow experiment.
 Since granular system tends to create collapsed solid phase, the shear 
motion normally occurs only within a mesoscopic layer 
 (several bead diameters wide)  near the moving interface.
Attempts to construct continuous description may be inadequate on
 this length scale. That is why 
 we  make $z$ a discreet variable, e.g. imagine
 that  the beads near the surface are organized in single-bead-wide layers 
(which in reality are reasonably well-defined). The layer at $z=0$ belongs
 to the moving wall, and has velocity $v_0$; in the limit $z=\infty$ all
 beads belong to a single coherent solid cluster and have zero velocity.

Consider $n$-th layer.  The typical rate with which a particle belonging to 
it is being hit by a particle at the adjoint layer, $n-1$, is $v_{n-1}/d$. Any 
time
 when this happens, there is a chance that the bead starts moving 
 (gets ``excited''). The particle will move by a certain 
distance $\delta$ (its  free path) until being  stopped by another particle at
 the same layer, as shown on Figure \ref{avalanche}(b) (one can easily modify the model to account 
for the possibility of "secondary excitations" to be born). 
As a result, 
the average velocity of the  $n-th$ layer is
\beq
v_n=\frac{\sigma (v_{n-1} +v_{n+1})}{d} \langle \delta\rangle_n
\label{vcouette1}
\eeq 
Here $\sigma$ is the coupling coefficient which measures the probability
 for a particle to be moved after being hit by a neighbor. We made sure 
that  $v_{n-1}$ and $v_{n+1}$ appear at the above equation in a
 symmetric manner. The asymmetry comes in 
boundary conditions:   once the
 position of the source of the excitations (the moving wall)  is specified, 
the velocities decay very rapidly with  $n$ and 
 one of the velocities   becomes negligible compared to the other. 

The obtained equation does not give a complete description of penetration of the 
motion inside the solid phase, because nothing was said about the typical 
inter-particle gap  $\langle \delta\rangle_n$.
It would be reasonable to assume that these gaps make a leading contribution to 
the free volume of the system, and therefore, $\langle \delta\rangle$ can be 
related to the deviation of the particle density $\rho_n$ from its 
random--closed--packing value $\rho_{cp}$:
\beq
\langle \delta\rangle_n\simeq  \frac {\rho_{cp}-\rho_n}{\rho_{cp}}d.
\label{delta}
\eeq 
Eqs. (\ref{vcouette1}) and (\ref{delta}) provide and opportunity to relate 
density and velocity profiles obtained experimentally. 
Note that constant $\langle \delta\rangle_n$ would result in exponential depth 
dependence of the velocity. It is more natural to expect the available free 
volume to decrease with depth $n$, 
thus resulting in a stronger than exponential velocity decay.

 We present  one possible way of modeling the  depth dependence 
of $\delta$. We shall call this treatment the  Quantized Free Volume approach.
It is based on the observation that after a particle jumps and gets stopped by 
its neighbors, the free volume available to it   becomes zero. 
As a result,  we expect the  free path  parameter $\delta$ to be
 either zero (for particles in contact with their front neighbors) 
 or of order of $d$ (for {\em movable} particles), as shown on Figure \ref{avalanche}(b). In the spirit of
 our earlier discussion, we interpret those movable particle as fluid
 component, and identify their local fraction with $\phi$. If 
 $\delta_*\sim d$ is a mean free path for the fluid  sub-system, then   

\beq
\langle \delta\rangle_n=\delta_*\phi_n.
\label{delphi} 
\eeq
 
\begin{figure}
 \centerline{ \epsfxsize=7.5truecm \epsfbox{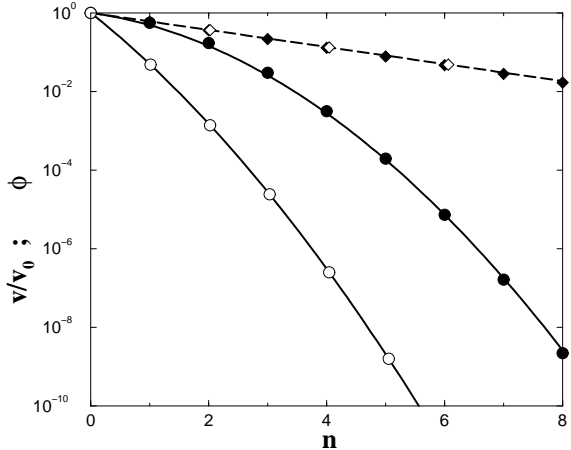}  }
\caption
{Velocity profiles (circles) and $\phi$-field (diamonds) in partially-fluidized system. White symbols correspond to the weak coupling regime ($\lambda_*=0.4$), and black ones correspond to the strong coupling ($\lambda_*=4.5$). One can clearly see the crossover from exponential to Gaussian velocity profile. Curves and data points represent analytic and numeric results, respectively. $\lambda=2$ for all curves.}
\label{expgauss}
\end{figure}

In order to find a missing equation for $\delta$, or $\phi$, we consider an
 exchange of free volume between adjoint layers. In the limit of $\phi \ll 1$, 
our assumption that the free volume is {\em quantized}, has an important 
implication. Since inter-layer free volume exchange may occur only near 
 movable particles (free volume quanta),  its flux is linear with their
 concentration. The  natural time scale of the volume  exchange between
 $n$-th and $n+1$-th layers is given by $v_{n-1}/d$ (the rate of change of
 the nearest neighborhood). Hence, the general expression for the
 corresponding free volume flux is 
\beq
\label{flux}
J_{n,n+1}=\frac{v_{n-1}}{d}(\sigma_{+}\phi_n-\sigma_{-}\phi_{n+1})+ 
\eeq
\[ 
\frac{v_{n+2}}{d}(\sigma_{-}\phi_n-\sigma_{+}\phi_{n+1}) 
\] 
Here $v_{n+2}$-term is added in order to make the structure of the
 model consistent with the inversion symmetry. In reality, after the 
boundary conditions are chosen,  one of the two terms can be neglected, since
velocity decays very rapidly with n.  In the case of shear flow, 
we can neglect $v_{n+2} \ll v_{n-1}$. 
In the steady state, the flux given by  (\ref{flux})  should vanish. The 
condition that $J_{n,n+1}=0$ provides the desired  closure for 
(\ref{vcouette1}-\ref{delphi}). The result is 
an exponentially-decaying  $\phi$--profile:
 \beq
\phi_n=\phi_0\exp\left(-\frac{n}{\lambda}\right). 
\label{phisol}
 \eeq
 Here $\lambda=1/(\log(\sigma_{-}/\sigma_{+}))$ is the penetration 
depth of the fluid component  into the solid bulk.   We assume  
$\sigma_{+}<\sigma_{-}$, since otherwise the solution (\ref{phisol}) 
 is clearly unphysical.  
Parameter $\phi_0\simeq 1$ depends on the detail of 
interactions between the moving wall an the first layer. Given the solution 
(\ref{phisol}), we can determine velocity field   from   (\ref{vcouette1})
\beq
v_n = V_0 \exp\left[-\frac{1}{2\lambda}\left(n+\frac{1}{2}+
\frac{\lambda}{\lambda_*}\right)^2\right]. 
\label{velsol} 
 \eeq
Here $\lambda_*=1/(\log(d/\delta_*\phi_0\sigma))$ is 
another length scale in the problem. Its variability from 
one system to another is expected mostly through the coupling coefficient
 $\sigma$. In the limit of very weak inter-layer coupling $\sigma$ 
(which corresponds to very smooth particles), 
$\lambda_*$ may become much less than $\lambda$, which would result 
in exponential velocity profile: $v_n\sim\exp(-n/\lambda_*)$. 
If the coupling is strong, both lengths $\lambda$ and $\lambda_*$ 
are expected to be of the same order, thus making
 the velocity profile  Gaussian--shaped. The analytic and numerical results of our model are presented  on Figure \ref{expgauss} both for strong and weak coupling regimes.
  
Existing  experiments do   show super-exponential  decay of the 
velocity profiles, in some systems it is remarkably close to Gaussian \cite{gauss,beringer}. Our model suggests the transition from 
exponential to Gaussian velocity profile depends on the strength of 
inter-layer coupling $\sigma$.  
Note that this  result differs from the conclusions of  
recent Debregeas--Josserand model \cite{GD.CJ},
 which attributes the change in velocity profile to the effect 
of dimensionality.

In conclusion, we have presented  a description of granular flows 
based on the notion that dissipation results in a strong correlation
 of particles' motion.   An extreme case of such correlated motion  is
  a coherent solid phase ($\phi=0$). We have constructed a model for the
 partial surface fluidization of this phase near a moving wall. The basic
 structure of our model employs  some concepts similar to those  of the
 conventional
 condensed matter theory: the motion is associated with  discreet 
excitations (movable particles) ``penetrating'' into the coherent 
solid phase within a mesoscopic surface layer. 
 We  made a quantitative
 prediction relating velocity and density profiles 
(\ref{phisol}-\ref{velsol}), 
which can be checked experimentally. One more check of our assumptions  
is possible by studying the velocity distribution function with high
 temporal resolution. If our approach is valid, one should  be able to 
identify the condensed and fluidized sub-systems with such kind of 
experiment.
In the case of 
 completely fuidized state, $\phi=1$, the collective behavior
 reveals itself through existence of short-living force chains. 
Once again, it  is  instructive to focus on    the  mesososcopic scales,
 set by the  
 typical length of a stable force chain segment (coherence length $\xi$). 
Our approach gives  several important results, which include $v(h)$ scaling 
 law for thick avalanches, as well as  velocity and density profiles 
 in granular  Couette--flow  experiment. Our conclusions are in a good
 agreement with existing experiment, and some of them are yet to be tested.

\section*{Acknowledgements} 
We thank  Georges Debregeas, Cristophe Josserand, Dan Mueth, Sidney Nagel, Heinrich Jaeger,  
Leo P. Kadanoff, Tomas A. Witten for fruitful discussions. 
This work was supported in part by  MRSEC Program of National Science Foundation under Award Number NSF DMR-9808595.

\end{document}